\documentclass[pra,twocolumn,aps,letterpaper, preprintnumbers,superscriptaddress]{revtex4-2}
\usepackage{graphicx}
\usepackage{amsmath,amssymb,amsfonts}
\usepackage{subfigure}
\usepackage{booktabs}

\usepackage{enumerate}
\usepackage{enumitem} 
\usepackage[usenames,dvipsnames]{color}
\usepackage[colorlinks=true,citecolor=blue,urlcolor=black]{hyperref}
\usepackage{ragged2e}
\usepackage{multirow}

\newcommand{\ket}[1]{\left\vert#1\right\rangle}

\begin{document}
	
	\title{Phase-Matching Quantum Key Distribution Without Intensity Modulation}
		
\author{Shan-Feng Shao}\thanks{The authors contributed equally to this work.}
\author{Xiao-Yu Cao}\thanks{The authors contributed equally to this work.}
\author{Yuan-Mei Xie}
\author{Jie Gu}
\author{Wen-Bo Liu}
\affiliation{National Laboratory of Solid State Microstructures and School of Physics, Collaborative Innovation Center of Advanced Microstructures, Nanjing University, Nanjing 210093, China}
\author{Yao Fu}\email{yfu@iphy.ac.cn}
\affiliation{Beijing National Laboratory for Condensed Matter Physics and Institute of Physics, Chinese Academy of Sciences, Beijing 100190, China}
\author{Hua-Lei Yin}\email{hlyin@nju.edu.cn}
\author{Zeng-Bing Chen}\email{zbchen@nju.edu.cn}
\affiliation{National Laboratory of Solid State Microstructures and School of Physics, Collaborative Innovation Center of Advanced Microstructures, Nanjing University, Nanjing 210093, China}

\date{\today}

\begin{abstract}
Quantum key distribution provides a promising solution for sharing secure keys between two distant parties with unconditional security. Nevertheless, quantum key distribution is still severely threatened by the imperfections of  devices. In particular, the classical pulse correlation threatens security when sending decoy states. To address this problem and simplify experimental requirements, we propose a phase-matching quantum key distribution protocol without intensity modulation. Instead of using decoy states, we propose a alternative method to estimate the theoretical upper bound on the phase error rate contributed by even-photon-number components. Simulation results show that the transmission distance of our protocol could reach 305 km in telecommunication fiber. Furthermore, we perform a proof-of-principle experiment to demonstrate the feasibility of our protocol, and the key rate reaches 22.5 bit/s under a 45-dB channel loss. Addressing the security loophole of pulse intensity correlation and replacing continuous random phase with six- or eight-slice random phase, our protocol provides a promising solution for constructing quantum networks.
\end{abstract}
\maketitle
	
	
\section{Introduction}
The fundamental principles of quantum mechanics open up endless and promising possibilities in fields such as communications, computing, and artificial intelligence~\cite{BENNETT20147,Ekert1991bell,Bennett1992bell,quantumcompu2019,quancompu2020pan,yin2022NSR,Quantummachinelearn,xie2021overcoming,zhou2022Research}. Quantum key distribution (QKD) is a tool for distributing secret keys between two remote parties, and it makes information-theoretic secure communication possible, even if the potential eavesdropper has unlimited computational power~\cite{BENNETT20147,Ekert1991bell}. Over the past decades, various protocols have been proposed for paving the way toward quantum networks~\cite{Bennett1992states,Bru1998six,Preskill2003secure,Kraus2005bound,renner2008security,koashi2009simple,Tomamichel2011uncertain,RevModPhys2012CV,Xu2020RMP,Liu2021CV}. Unfortunately, there are many security loopholes in QKD caused by the imperfections of experimental devices~\cite{Xu_2010,hacking2010,sourceattack2013pan,xu2015experimental,pereira2019quantum,lasatt2019huang,imperf2023huang}.
Measurement-device-independent (MDI) QKD removes all the side channels of the measurement unit~\cite{Lo2012MDI}. Thus far, many theoretical and experimental breakthroughs have been made in MDI QKD~\cite{Lo2012MDI,vulnerabilities2016Comandar,Hualei2016404km, Asymme2019Lo,Cao2020LongMDI,fan2021MDI,gujie2022exper,fan2022robust}. Twin-field QKD~\cite{TF-QKD2018Lucamarini}, a variant of MDI QKD, which uses single-photon interference, has triggered many works~\cite{ma2018phase,SNS2018Wang, yin2019measurement,hu2019SNS,Curty2019simp,yin2019coher,PhysRevLettTF2019,maeda2019repeaterless,wang2020opt,zeng2020sym,Li:21,FangZeng2020,PittalugaMinder2021,cz2021,WangYin2022,zhou2022,xie2023Scalable} to break the rate-loss limit~\cite{pirandola2017fundamental}. By using the postdetection event pairing, asynchronous MDI QKD or called mode-paring QKD has been recently proposed~\cite{xie2022RPX,zeng2022mode} and experimentally demonstrated~\cite{zhou2022experimental,zhu2023experimental} to allow repeaterlike rate-loss scaling. Surprising, the asynchronous MDI QKD has a higher key-rate advantage in the intercity range~\cite{Xie2023Advantages,bai2023asynchronous}.
	
However, the security of most current QKD protocols relies on accurate modulation of the optical intensity. The decoy-state method~\cite{global2003Hwang,beat2005wang,Lo2005decoy} usually utilizes pulses with different intensities to estimate the bounds on phase error rate in postprocessing. Although there are a large number of works applying this method for better estimation~\cite{PNS2000Brass,PNS2000Lutk}, the correlation between different pulse intensities becomes another security issue~\cite{Tamaki_2016,Akihisa2018correl,Nagamatsu2016dependently,zapatero2021security,PhysRevApplied.18.044069,wang2022fully}. The deviation of the intensity is the most obvious phenomenon because of the classical pulse correlation~\cite{gujie2022exper}, and the deviation will leak a lot of information to eavesdroppers. Solving the issue of correlation in intensity modulation has led to the development of ingenious, yet complex, methods for proving security~\cite{Akihisa2018correl,zapatero2021security,PhysRevApplied.18.044069,wang2022fully}. However, it is worth noting that these approaches often come with significantly reduced secret key rates and require intricate experimental setups. 
	
Here, we present a phase-matching QKD protocol that avoids the pulse correlation problem caused by intensity modulation and provide the measurement-device-independent characteristic~\cite{sidechannel2012,threeDI2013wang, threeMDI2013wang, Deatt2013Lucio,yin2014long,finiMDI2014Lo,PhaseRF2015Guo,Wang17MDIenvir,Highrate2015Andersen, statisMDI2015wang, zhou2016making}. Our protocol does not require intensity modulation, providing a more robust approach for QKD. Besides, in the phase-matching QKD, phase randomization is needed (typically 16-slice random phase), and the even-photon-number states contribute to the whole phase error rate~\cite{yin2019coher,zeng2020sym}. The key innovation of our work lies in the utilization of a alternative estimation method to obtain the theoretical upper bound on the phase error rate without the need for intensity modulation To obtain the upper bound on the vacuum state phase error rate, we assume that the quantum bit error rate (QBER)  only comes from the vacuum state. Based on the new estimation method, we need only six or eight-slice random phases due to the relatively low pulse intensity. Without the need of modulating decoy state and vacuum state, the experimental operation is simplified. To demonstrate the feasibility of our protocol, we also perform a proof-of-principle experiment, and achieve a key rate of 22.5 bit/s under a 45-dB channel loss. This verifies the potential of our protocol for general application scenarios.

\section{Protocol Description}
 
\begin{figure}[b]
\centering
\includegraphics[width=1\columnwidth]{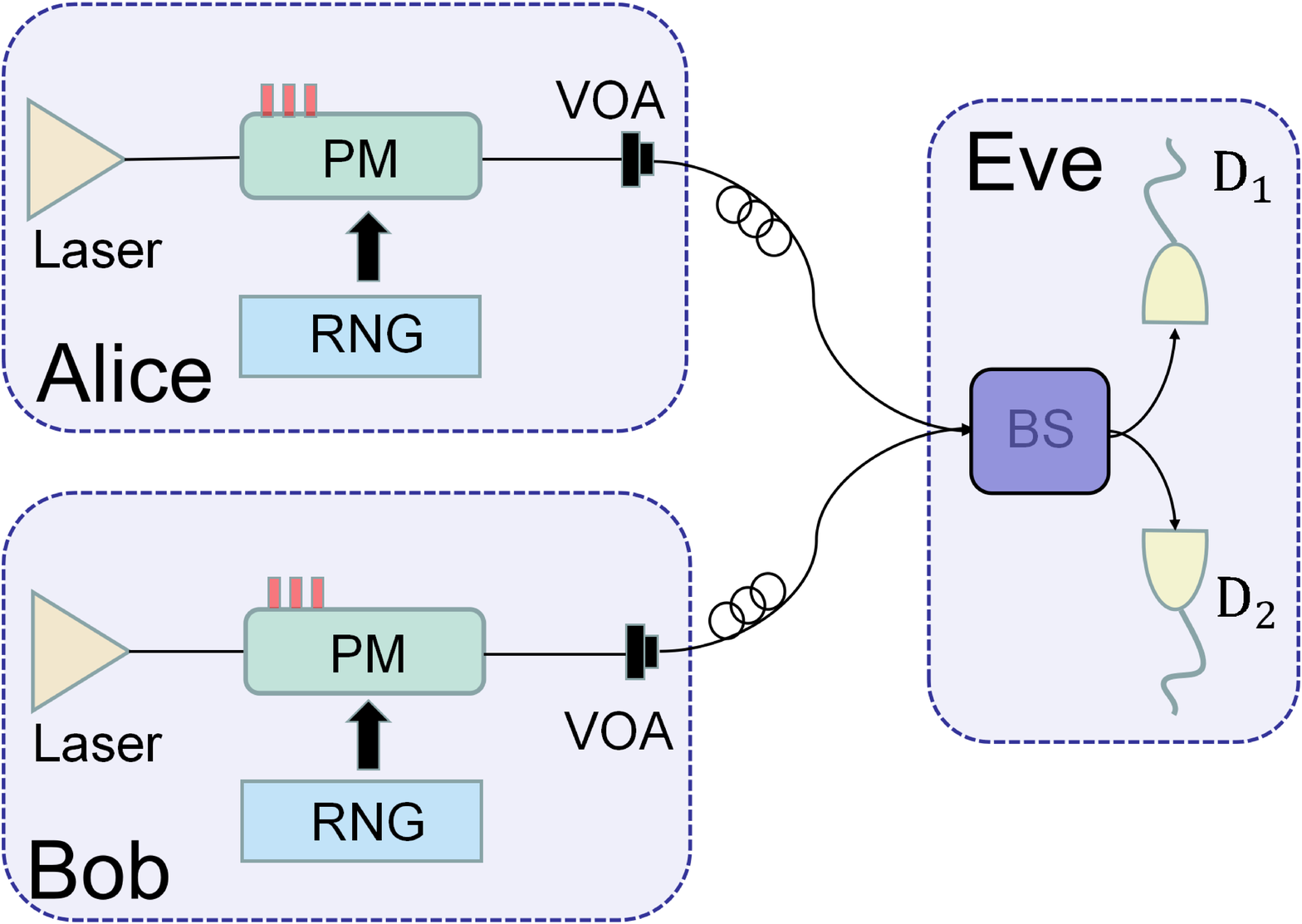}
\caption{Schematic of phase-matching QKD without intensity modulation. Alice and Bob utilize pulse laser sources to prepare weak coherent states. They use a random number generator (RNG) to generate random numbers for random phases and random key bits. For each pulse train, Alice and Bob exploit a phase modulator (PM) to apply phase $\theta_{a} + r_{a}\pi$ and $\theta_{b}+ r_{b}\pi$ on each pulse according to the random phase selection and random key bits $r_{a}$ and $r_{b} \in \{0, 1\}$, respectively. A variable optical attenuator (VOA) is used to implement a weak pulse with single-photon-level modulation. An untrusted Eve receives the two pulses from Alice and Bob and then uses a beam splitter (BS) and single-photon detectors to conduct an interference measurement.} \label{setup}
\end{figure}

A schematic of our protocol is shown in Fig.~\ref{setup}. In our protocol, Alice and Bob each generate weak coherent states independently and apply respective phases to them. These modulated pulses are then transmitted to Eve, who performs an interference measurement. Eve declares a valid click only when a single detector registers a click. The details of our protocol are given below. 
 
~\noindent{\it{1.~Preparation.}} Alice and Bob independently prepare weak coherent states $| \sqrt{\mu_{a}} e^{i(\theta_{a}+r_{a}\pi)}\rangle~$ and $| \sqrt{\mu_{b}} e^{i(\theta_{b}+r_{b}\pi)}\rangle~$ and send them to an untrusted party, Eve. $r_{a}, r_{b} \in \{0,1\}$ are random key bits;  $\theta_{a}, \theta_{b}\in \left\{\frac{2\pi}{M},2\frac{2\pi}{M},...,M\frac{2\pi}{M}\right\}$ are globally random phases. $M$ is the number of random phase slices. $\mu_{a}$ and $\mu_{b}$ are the pulse intensities of Alice and Bob, respectively. $\mu_{a}+\mu_{b}=\mu$ is the total pulse intensity.

~\noindent{\it{2.~Measurement.}} Eve uses the two pulses from Alice and Bob to conduct an interference measurement and chooses a single detector ($D_{1}$ or $D_{2}$) click as a valid click. 
	
~\noindent{\it{3.~Sifting.}} After measurement, Eve announces the clicking detector when a valid click occurs. Then, Alice and Bob announce their corresponding random phases. They will keep the data if $|\theta_{a}-\theta_{b}|$ = 0 or $\pi$. If detector $D_{2}$ clicks and $|\theta_{a}-\theta_{b}|$ = 0 (if $D_{1}$ clicks and $|\theta_{a}-\theta_{b}|$ = $\pi$), Bob will flip his bit. Steps 1 to 3 are repeated $N$ times until the data is sufficient to conduct the steps below.
	
~\noindent{\it{4.~Parameter estimation.}} Alice randomly samples some data with probability $p_{s}$ as the test data and announces the locations and bits information. Bob calculates the bit error number $m_{s}$ of test data and announces to Alice. The rest of the data serve as the shift key.

\begin{figure*}[t]
\centering
\includegraphics[width=16 cm]{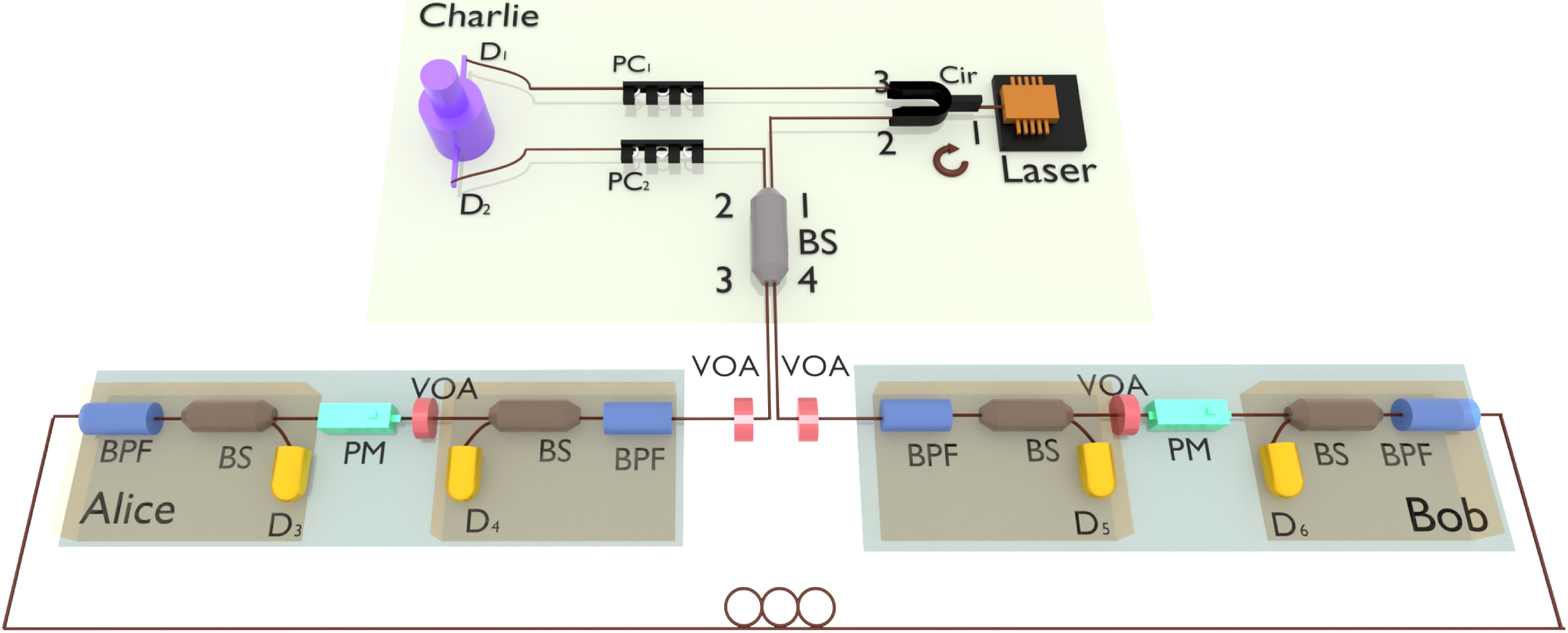}
\caption{Schematic of the proof-of-principle experiment. Detectors $D_{1}$ and $D_{2}$ are superconductor nanowire single-photon detectors. Detectors $D_{3}$, $D_{4}$, $D_{5}$, and $D_{6}$, which monitor the intensity of pulses, are photodiodes. All the devices are synchronized by the signals from an arbitrary waveform generator. The pulses are emitted from a homemade laser and reach a circulator (Cir) whose direction is from 1 to 2 and from 2 to 3.} Then, the pulse is separated into two pulses by a 50:50 BS, and the two pulses enter a Sagnac loop. Before modulation, the pulse passes through a bandpass filter (BPF) and the BS. Alice and Bob can distinguish whether a pulse belongs to them after time calibration. The two pulses modulated by Alice and Bob passing through the Sagnac loop will interfere with each other by the BS. Polarization controllers (PCs) modify the polarization of the two pulses to maximize the detection efficiencies. Finally, the pulses after interference are then detected by two single-photon detectors $D_1$ and $D_2$. Note that the devices covered with the yellow cuboid are not introduced in the implementation.\label{experiment928}
\end{figure*} 	
  
~\noindent{\it{5.~Postprocessing.}} Finally, Alice and Bob conduct error correction and privacy amplification. After that, Alice and Bob obtain the final secret keys. 

 The differences between our protocol and conventional phase-matching QKD~\cite{ma2018phase} can be summarized as follows. Firstly, our protocol eliminates the requirement for intensity modulation, thereby avoiding the pattern effect associated with it. Secondly, our protocol utilizes fewer phase slices compared to conventional phase-matching QKD due to the lower intensity.
   
\section{Experimental Demonstration}	
		
To demonstrate the feasibility of our protocol, we implement a proof-of-principle experiment. The experimental setup is shown in Fig.~\ref{experiment928}. We exploit a Sagnac loop to stabilize the fluctuation of the phase caused by the path~\cite{sagnac2019zhong}.

The laser source is held by the third party, Charlie. The frequency of the pulse laser is set as 100 MHz, and the duty cycle is less than $3\%$. Charlie utilizes the laser to generate pulses sent to Alice and Bob. The pulses pass through a circulator (Cir) and a 50:50 BS whose port numbers are shown in Fig.~\ref{experiment928}. Then, the two identical pulses will enter the loop. Alice and Bob capture and modulate their own pulses. The rule of modulating the corresponding pulses is as following: Alice modulates the clockwise pulse, and Bob modulates the counterclockwise pulse. In the loop, we utilize four BPFs for filtering and four BSs and detectors to monitor the injected pulse intensities. We did not realize pulse filtering and intensity monitoring because of the device limitations. The impact on the results caused by the lacking devices is too slight to consider. As mentioned above, different random phases are generated with an equal probability 12.5\%, and key are selected with a probability 50\%. Therefore, we used Python to generate a set of random numbers for the arbitrary waveform generator (Tabor Electronics, P2588B). In our implementation, we select eight slices of the random phase, which is more complex than the six slices in experiment, and the length of the random number is 10000. Then, the radio-frequency signals are amplified by an electrical driver to drive the PM to modulate the total phase. The pulses modulated by Alice and Bob pass through the VOA and different BS ports to the detection units. After interference at the BS, the pulses are detected by $D_1$ and $D_2$. For $D_1$, the detection efficiency is 86.3\% and the dark count rate is 13.1 Hz. For $D_2$, the detection efficiency is 82.9\% and the dark count rate is 18.9 Hz. The losses of Charlie’s components are presented in Table~\ref{EFF_OPT} at the end. The length of the time window is 1.8 ns. We ran the system for 1000 seconds under different channel losses to accumulate sufficient detection events and distilled the raw keys.
		
\section{Security Analysis}
	
For phase-matching QKD~\cite{ma2018phase}, the phase error rate is only related to the even-photon component~\cite{yin2019coher,zeng2020sym}. Strictly, for security proofs based on photon-number states, continuous phase randomization is required. We use a discrete phase randomization ($M=6, 8$) to replace continuous phase randomization to enhance the accessibility. Initially, we calculate the phase error rate for the case of continuous  phase randomization. Then, we analyze the deviation that occurs in our protocol when employing discrete modulations. Based on the analysis, the discrete random phase can be used to replace the continuous random phase even if $M$ is 6 or 8. 

\subsection{Continuous-phase randomization}
 
After the continuous-phase randomization, the joint system between Alice's and Bob's can be regarded as a mixture of photon-number states
\begin{equation}
\begin{aligned}
\rho_{ab}&=	\frac{1}{2\pi} \int_{0}^{2\pi} |\mu e^{i\theta}\rangle_{ab}\langle\mu e^{i\theta}| d\theta 
\\&= \sum_{k=0}^{\infty}P_{k}|k\rangle_{ab}\langle k|,
\end{aligned}
\end{equation}
where we have the probability of joint $k$-photon $P_{k}=e^{-\mu}\mu^{k}/k!$ and $k$ is the joint photon number between Alice and Bob. The phase error rate can be written as~\cite{yin2019coher,zeng2020sym}
\begin{equation}
\begin{aligned}
E_{p}=\sum_{k=0}^{\infty}q_{2k}=\frac{1}{Q_{\mu}}\sum_{k=0}^{\infty}P_{2k}Y_{2k},\label{pe}
\end{aligned}
\end{equation}
where $q_{k}=P_{k}Y_{k}/Q_{\mu}$ is the ratio of joint $k$-photon in the final valid detection event. $Q_{\mu}$ is the gain of 
Alice and Bob sends optical pulses with intensities $\mu_{a}$ and $\mu_{b}=\mu-\mu_{a}$, respectively. $Y_{k}$ is the yield of joint $k$-photon between Alice and Bob (
$0\leq Y_{k}\leq 1$).

To estimate the yield of vacuum state $Y_{0}$, one needs to randomly sample some bits to obtain the QBER. The observed value of the sampled bit error number is $m_{s}$. Then, we use the variant of the Chernoff bound~\cite{yin2020tight} to estimate the expected upper bound of the sampled bit error number $\overline{m}_{s}^{*}=\phi^{U}(m_{s})$, where $\phi^{U}(x) =x+\beta+\sqrt{2\beta x+\beta^{2}}$, $\beta = \rm{log}(\epsilon^{-1})$ and $\epsilon$ is the failure probability. Therefore, 
the upper bound of the expected bit error number $m^{*}$ in the shift key can be given by 
\begin{equation}
\begin{aligned}
\overline{m}^{*}=\frac{(1-p_{s})}{p_{s}}\overline{m}_{s}^{*}.
\end{aligned}
\end{equation} 
The expected value of error data number $\overline{m}_{0}^{*}$ caused by vacuum state is not greater than the total error data number $\overline{m}^{*}$, namely, $\overline{m}_{0}^{*}\leq\overline{m}^{*}$. A useful observation is that zero photon will result in half the expected error detection data, i.e., $\overline{n}_{0}^{*}=2\overline{m}_{0}^{*}$ and $n_{0}^{*}$ is the expected value of vacuum state's contribution.
Therefore, the upper bound of $Y_{0}$ can be given by
\begin{equation}
\begin{aligned}
\overline{Y}_{0} = \frac{\overline{n}_{0}}{N(1-p_{s})e^{-\mu}},\label{Y0}
\end{aligned}
\end{equation}
where the observed value $\overline{n}_{0}=\Phi(\overline{n}_{0}^{*})$ is calculated by the Chernoff bound~\cite{yin2020tight} $\Phi^{U}(x) =x+\beta/2+\sqrt{2\beta x+\beta^{2}/4}$.

Combining the discussion above, we incorporate the upper bound of observed value of $\overline{Y}_{0}$ and probability distribution $P_{2k}$ into the formula of phase error rate

\begin{equation}
\begin{aligned}
E_{p} &= \frac{1}{Q_{\mu}}P_{0}Y_{0} + \frac{1}{Q_{\mu}}\sum_{k=1}^{\infty}P_{2k}Y_{2k}\\
&\leq \frac{1}{Q_{\mu}}P_{0}Y_{0} + \frac{1}{Q_{\mu}}\sum_{k=1}^{\infty}P_{2k}\\
&\leq \frac{e^{-\mu}\overline{Y}_{0}}{Q_{\mu}}+\frac{e^{-2\mu}+1-2e^{-\mu}}{2Q_{\mu}}.\label{Ep}
\end{aligned}
\end{equation}
For the purpose of estimating the upper bound of the phase error rate, we set the worst case that $Y_{2k}=1$ with $k\geq 1$ considering the negligible $P_{2k}$ at the second line of Eq.~\eqref{Ep}.

\subsection{Discrete-phase randomization}

For the case of discrete random phase modulation, the system will become a group of ``pseudo" Fock states according to the density matrix of the states that Alice and Bob prepare \cite{Cao_2015}
\begin{equation}
		\frac{1}{M}  \sum_{j=0}^{M-1}|\sqrt{\mu}e^{i\theta_{j}}\rangle\langle \sqrt{\mu}e^{i\theta_{j}}| = \sum_{k=0}^{M-1} P^{\mu}_{M}(k)|\lambda_{k}\rangle\langle \lambda_{k}|,
		\end{equation}
		where
		\begin{equation}
		|\lambda_{k}\rangle=\frac{e^{-\mu/2}}{\sqrt{P^{\mu}_{M}(k)}}\sum_{l=0}^{\infty}\frac{(\sqrt{\mu})^{lM+k}}{\sqrt{(lM+k)!}} |lM+k\rangle,
		\end{equation}
    and
		\begin{equation}
		P^{\mu}_{M}(k)=\sum_{l=0}^{\infty}\frac{(\mu)^{lM+k}e^{-\mu}}{(lM+k)! }.
		\end{equation}
Observing this form, the state becomes the Fock state when the $M$ is large enough. If $M$ is even, each pesudo even Fock state $\ket{\lambda_{k}}$ contains only even photon-number states. Based on the phase-matching QKD protocol analysis~\cite{yin2019coher,zeng2020sym}, the phase error rate is contributed only by the even-photon components. Therefore, we can consider only the deviation of even-photon component when the random phase slices $M=8$. The pesudo even photon numbers are $\{0, 2, 4, 6\}$. Furthermore, we test the $M=6$ and the pesudo even numbers are taken as $\{0, 2, 4\}$. The even-photon deviation is shown below~\cite{zeng2020sym}, and more details are presented in Appendix~\ref{even}
\begin{equation}
\begin{aligned}
\delta_{k} = |q_{k}-q_{k}^{M}|
  \leq \frac{P^{\mu}_{M}(k)} {Q_{\mu}}\sqrt{\frac{k!\mu^{M}}{(M+k)!}}.\label{devia}
\end{aligned}
\end{equation}
From Eq.~\eqref{pe}, the deviation will cause the extra phase error rate. We could write the total phase error rate with $M$-slice random phase
\begin{equation}
\begin{aligned}
E_{p}^{M} & =\sum_{k=0}^{M/2-1} q_{2k}^{M} \\
 &\leq \frac{e^{-\mu}\overline{Y}_{0}}
 {Q_{\mu}}+\frac{e^{-2\mu}+1-2e^{-\mu}}{2Q_{\mu}}+\sum_{k=0}^{M/2-1}\delta_{2k}. \label{devia2}
\end{aligned}
\end{equation}
The detailed derivation of this problem has been carried out in Appendix~\ref{discrete}.
Furthermore, we use the Kato inequality~\cite{kato2020concentration} to defend against the coherent attacks for the dependent random variables. Further details regarding this approach are provided in Appendix~\ref{Kato}.

\section{Simulation Results}
 
Let us define  $\xi'$ as the bits consumed to ensure that the failure probability of error verification reaches $2^{-\xi'}$, and  $\xi$ denotes the additional amount of privacy amplification to further enhance the privacy. According to complementarity~\cite{koashi2009simple,maeda2019repeaterless}, an 
$\epsilon_{\rm sec}$ -secret and $\epsilon_{\rm cor}$-correct key of length is

\begin{equation} 
\ell = n_{\mu}[1-H(\overline{E}_{p}^{M})-fH(E_{b})]- \xi -  \xi' ,
\end{equation}
where $n_{\mu} = \frac{2}{M}Q_{\mu}N(1-p_{s})$ is the remaining bit number to generate the logic bits, $H(x) = - x{\rm log_{2}}x - (1-x){\rm log_{2}}(1-x)$ is the  Shannon entropy function, $E_{b}=e_{d}(1-p_{d})[1-(1-p_{d})e^{-\mu\eta}]/Q_{\mu}+(1-e_{d})p_{d}(1-p_{d})e^{-\mu\eta}/Q_{\mu}$ is the QBER, and $\overline{E}_{p}^{M}$ is the total phase error rate with phase slice number $M$ after applying  Kato inequality to defend against the coherent attacks. The total gain $Q_{\mu}=(1-p_{d})[1-(1-2p_{d})e^{-\mu\eta}]$.
	\begin{table}[t]
		\caption{Simulation parameters~\cite{2020PRLeffi56}. The misalignment error and dark count rate are denoted by $e_{d}$ and $p_{d}$, respectively. $f$ is the error correction efficiency, $\eta_{d}$ is the detector efficiency, $\alpha$ is the loss coefficient, and $M$ is the number of phase slices whose value will influence the even-photon deviation. }
  		\setlength{\tabcolsep}{3.8mm}

			\begin{tabular}{cccccc}
				
			\hline
				\hline

				 $e_{d}$&$p_{d}$&$f$ & $\eta_{d}$&$\alpha$&$M$\\
				 \hline
				0.01 & $10^{-8}$ &1.16&56\%&0.168&6 or 8\\

			\hline
				\hline
				\label{parameter}
			\end{tabular}
		\end{table}
  
Due to the application of the Chernoff bound twice, the security parameters can be expressed as $\epsilon_{\rm sec}=\sqrt{2}\sqrt{2\epsilon+2^{-\xi}}$ and $\epsilon_{\rm cor}=2^{-\xi'}$. Furthermore, we utilize the Kato inequality to defend against the coherent attacks for the dependent random variables. The $\epsilon_{\rm{Ka}}$ is the failure probability in Kato inequality. Therefore, the final
security parameter is $\epsilon_{\rm tot}=\epsilon_{\rm sec}+\epsilon_{\rm cor}+\epsilon_{\rm{Ka}}=\sqrt{2}\sqrt{2\epsilon+2^{-\xi}}+2^{-\xi'}+\epsilon_{\rm{Ka}}$.
In the simulation, we set $\epsilon_{\rm sec}=2\times10^{-10}$, $\epsilon_{\rm cor}=10^{-15}$ and $\epsilon_{\rm{Ka}}=10^{-10}$. We can conclude that  $\epsilon=10^{-20}/2$, $\xi=\rm{log}_{2}(2/10^{-20})$,  $\xi'=\rm{log}_{2}(1/10^{-15})$ and $\epsilon_{\rm tot}= 3\times 10^{-10}$.
\begin{figure}[b]
\centering
\includegraphics[width=8.6 cm]{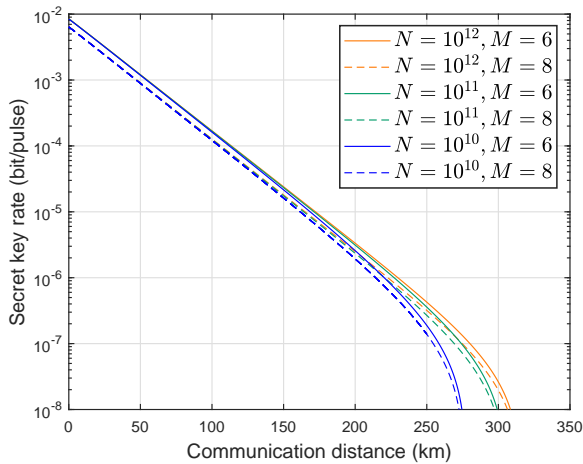}
\caption{Finite-size simulation results of our protocol. The parameters are shown in Table~\ref{parameter}. We select data sizes $N = 10^{10}$, $ 10^{11}$, and $10^{12}$ to conduct the simulation.} \label{finite}
\end{figure}

Here, we numerically simulate the key rate $R = \ell/N$ of our  protocol in finite-size cases. The other parameter settings are shown in Table~\ref{parameter}. The finite-size simulation results are shown in Fig.~\ref{finite}. 
 
Our protocol achieves a 305-km transmission distance with $N = 10^{12}$. At the condition of $N=10^{11}$, the transmission distance can reach 298 km with the $10^{-8}$ key rate. Even when the data size is not a large number, such as $10^{10}$, the transmission distance reaches 270 km. 

\begin{figure}[t]
\centering
\includegraphics[width=8.6 cm]{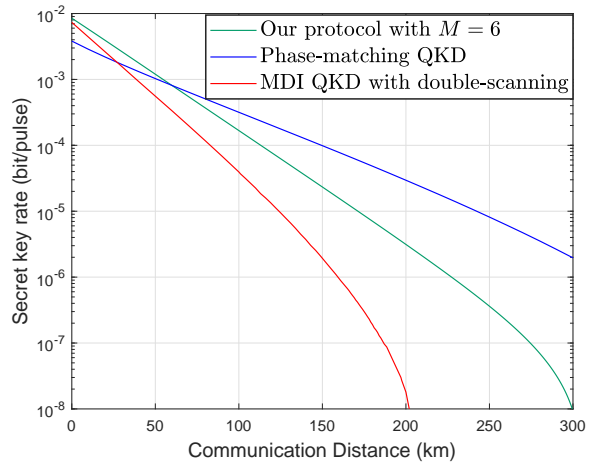}
\caption{We compare the performance of our protocol with phase-matching QKD~\cite{ma2018phase} and four-intensity MDI QKD using the double-scanning method~\cite{jiang2021higher}. The simulation parameters are provided in Table~\ref{parameter} with a data size of $N = 10^{11}$. The phase slice number of our protocol $M$ is set as 6 in comparison. The total security parameter $\epsilon_{\rm{tot}}$ of four-intensity MDI QKD using the double-scanning method is the same for that of our protocol, which is equal to $3\times 10^{-10}$.
 } \label{compare}
\end{figure}

From Fig.~\ref{compare}, we can observe that our protocol demonstrates superior performance within a 60-km range, and the key rate achieved by our protocol is comparable to that of phase-matching QKD at a distance of 100 km. This implies that our protocol achieves a similar key rate to phase-matching QKD in metropolitan areas without the need for intensity modulation. Compared to four-intensity MDI QKD with double-scanning method, our protocol demonstrates a key rate advantage of approximately 4 times at a distance of 100 km.
  
QKD focuses on applicability and security rather than the only transmission distance. In the case of intercity communication where the distance is 500 km or more, the key rate is too low to be practically applicable. However, within metropolitan areas, such as in 100-km distance communication which is the main field of application. In such a situation, our protocol achieves a key rate comparable to that of phase-matching QKD while overcoming the pattern effect and simplifying experimental requirements. It is worthwhile to note that phase-matching QKD and four-intensity MDI QKD using the double-scanning method require perfect intensity modulation, which is impractical in experiments. Additionally, the pattern effect in these methods leaks information to potential eavesdroppers. These issues~\cite{PhysRevApplied.18.044069,Akihisa2018correl} significantly decrease the key rate of phase-matching QKD and four-intensity MDI QKD with double-scanning method, further highlighting the competitiveness of our protocol.

\begin{figure}[t]
    \centering
    \includegraphics[width=8.6 cm]{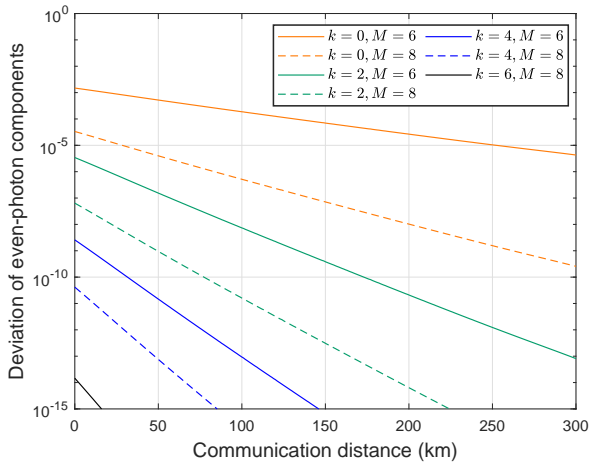}
    \caption{Deviation of the even-photon state in our protocol. We simulate the deviation with the data size $N=10^{11}$. The other parameters chosen are shown in Table~\ref{parameter}. The deviation of the even-photon state can denote the gap between the continuous random phase and discrete random phase.
    } \label{deviation}
\end{figure}

The change in the deviation with attenuation is shown in Fig.~\ref{deviation}.
We find that the influence of the deviation is too small to consider. From Eq.~\eqref{devia2}, the deviation has such a negligible effect on the phase error rate that it can be disregarded, and the phase error rate increases by less than $1\%$ of itself when using 6-slice random phase. The substitution of fewer slices does not invalidate the security proof that relies on photon-number states. In some extreme circumstances, such as with a high source intensity, we have to utilize more slices to achieve the replacement, but this comes at the expense of some key rate and experimental complexity. Because the intensity of the pulse we use is sufficiently low, we can use a small number of phase slices, which is set as 6 or 8 in our protocol, to replace the continuous random phase. 
	\section{Experimental Results}

		\begin{table*}[t]
			\center
			\caption{Summary of experimental data. We tested the key rate under different channel losses. The table shows the data size $N$, the total intensity of pulses $\mu$, experimental QBER $E_{b}$, number of the remaining bits $n_{\mu}$, and key rate $R$. Note that losses of 35, 40, and 45 dB correspond to $p_s$ of $7\%$.}
   			\setlength{\tabcolsep}{0.9cm} 
      \renewcommand\arraystretch{1.5}
			\begin{tabular}
				{cccccc}  \hline \hline
				Loss & $N$ & $\mu$ & $E_b$ &  $n_{\mu}$&  $R$ \\ \hline
				35 dB & $10^{11}$ & 3.20$\times 10^{-3}$ &0.22\% & 934403 & 3.00$\times 10^{-6}$ \\ 
				40 dB & $10^{11}$ &1.87$\times 10^{-3}$ & 0.33\%  &302187 & 8.50$\times 10^{-7}$\\
                45 dB & $10^{11}$ & 9.78$\times 10^{-4}$ &0.71\%  &91781 & 2.25$\times 10^{-7}$\\
                            \hline \hline
			\end{tabular}
			
			\label{result_table}
		\end{table*}

\begin{figure}[tb]
			\centering
			\includegraphics[width=86mm]{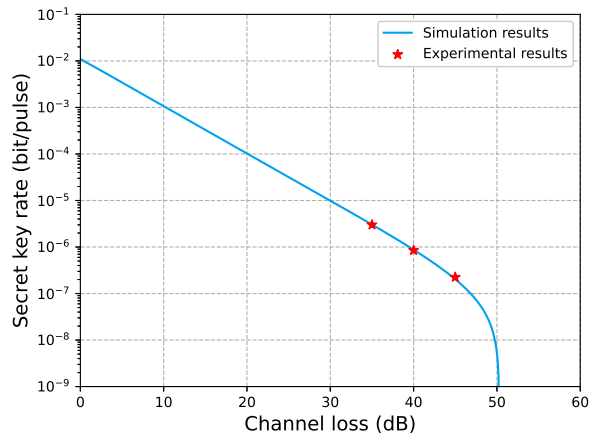}
			\caption{ Considering the experimental results, we depict the secret key rate when the data size  $N = 10^{11}$ and phase slice $M=8$. We implement the experiment with the optimized intensity under 35 dB, 40 dB and 45 dB channel losses.}
			\label{rate_exp}
\end{figure}

We implement a proof-of-principle experiment to test our protocol under 35-, 40-, and 45-dB channel losses with the experimental setup depicted in Fig.~\ref{experiment928}. The experimental results we obtained are listed in Table~\ref{result_table} and Fig.~\ref{rate_exp}.

We implement experiments and obtain the total detection counts and total QBER $E_{b}$ under the total intensity $\mu$ 
with $N = 10^{11}$. The optimized pulse intensity is acquired by using the genetic algorithm in simulations with different channel losses. Given the 100-MHz repetition rate, our protocol can obtain a secure key rate of 22.5 bit/s when the channel loss is over 45 dB, which means that it can be implemented over 267 km with existing technologies. A secure key rate of 0.3 kbit/s is generated at 35 dB ($\sim$208 km), while at 40 dB ($\sim$238 km), the rate is 85 bit/s. The more details of experiment is shown in Appendix~\ref{exper details}.
As a proof-of-principle demonstration, the aim of implementing the scheme is to verify the feasibility of our protocol instead of establishing a complete system. We used the Sagnac loop to stabilize the phase automatically, and thus, the pulses modulated by the two users were generated by a third party, which resulted in security flaws. For real optical fiber implementation, our protocol can be performed by using the phase locking and phase tracking method to replace the Sagnac loop, where two users are independent.

\section{Conclusion}
In this work, we propose a phase-matching QKD protocol without intensity modulation. Since the need of sending decoy states to calculate the secret key rate is removed, our protocol avoids the pulse correlation resulting from multi-intensity modulation and simplifies the experimental requirements. A alternative estimation method is introduced to obtain the phase error rate in our security analysis, which assumes that the vacuum state contributes to the whole QBER in order to obtain an upper bound on the vacuum state phase error rate. 

We utilize the Kato inequality to defend against the coherent attacks for the dependent random variables. Based on these considerations, we conducted a simulation to demonstrate the key rate of our protocol. Simulation results show that our protocol can reach 305 km with a data size of $N=10^{12}$. Our protocol demonstrates a advantage of approximately 4 times the magnitude in the key rate compared to the four-intensity MDI QKD with the double-scanning method at 100 km. Furthermore, within metropolitan areas, our protocol achieves a comparable key rate to that of phase-matching QKD. Considering the imperfect intensity modulation and its pattern effect, our protocol becomes even more competitive. The feasibility of using discrete random phase with fewer phase slices (M = 6, 8) has been demonstrated by the simulation results. With low pulse intensity, we need only a six-slice or eight-slice random phases, which further reduces the experimental complexity and saves the random number resources. A proof-of-principle experiment is implemented to demonstrate the feasibility of our protocol, and the experimental results shows that our protocol can achieve a key rate of 22.5 bit/s under a 45 dB channel loss.  The experimental results are consistent with the simulation results. The simplicity and efficiency of our protocol, achieved through the avoidance of intensity modulation and the use of fewer slice random phases, make it a practical solution for quantum communication.

		\section*{Acknowledgements}
		This study was supported by the National Natural Science Foundation of China (No. 12274223), the Natural Science Foundation of Jiangsu Province (No. BK20211145), the Fundamental Research Funds for the Central Universities (No. 020414380182), the Key Research and Development Program of Nanjing Jiangbei New Area (No. ZDYD20210101),  the Program for Innovative Talents and Entrepreneurs in Jiangsu (JSSCRC2021484), and the Program of Song Shan Laboratory (included in the management of the Major  Science and Technology Program of Henan Province) (No. 221100210800-02).
		
\appendix

		\section{Deviation of even-photon components}
  \setcounter{equation}{0}
\renewcommand{\theequation}{A\arabic{equation}}
   \label{even}
  The following derivation is based on Ref. \cite{zeng2020sym}. Note that $P^{\mu}_{M}(k) \geq P_{k}$, the deviation of even-photon components can be bounded by
  		\begin{equation}
		\begin{aligned}\label{A1}
		|q_{k}-q_{k}^{M}| & = \frac{|Y_{k}P_{k}-Y_{k}^{M} P^{\mu}_{M}(k)|}{Q_{\mu}}
  \\& \leq P^{\mu}_{M}(k) \frac{|Y_{k}-Y_{k}^{M}|}{Q_{\mu}},  
		\end{aligned}
		\end{equation}    
  where $Y_{k}^{M}$ is the yield of joint $k$-photon with $M$-slice random phase. The deviation of yield is bounded with $|Y_{k}-Y_{k}^{M}| \leq \sqrt{1-|\langle k|\lambda_{k}\rangle|^{2}}$. Further, we get
    \begin{equation}
		\begin{aligned}
		|\langle k|\lambda_{k}\rangle|^{2} &= \frac{e^{-\mu}}{P^{\mu}_{M}(k)} \left | \sum_{l=0}^{\infty} \frac{(\sqrt{\mu})^{lM+k}}{\sqrt{(lM+k)!}}\langle k|lM+k\rangle \right |^{2}
  \\&=\frac{e^{-\mu} \mu^{k}}{P^{\mu}_{M}(k) k!}
   \\&=\frac{1}{ \left (\sum_{l=0}^{\infty} \frac{k!}{
(lM+k)!}\mu^{lM}\right)}.
		\end{aligned}
		\end{equation}  
  Here, we take an inequality $[(l+1)M+k]!\geq[(lM)+k]!(M+k)!/k!$ into the formula above
    		\begin{equation}
		\begin{aligned}
  |\langle k|\lambda_{k}\rangle|^{2} &\geq \frac{1}{\sum_{l=0}^{\infty}\left(\frac{k!}{(M+k)!}\mu^{M}\right)^{l}}
   \\&=1-\frac{k!}{(M+k)!}\mu^{M}.
		\end{aligned}
		\end{equation}   
  Then, we take the formula into the bound of even-photon deviation 
  		\begin{equation}
		\begin{aligned}
		|q_{k}-q_{k}^{M}| &\leq \frac{P^{\mu}_{M}(k)} {Q_{\mu}}\sqrt{\frac{k!\mu^{M}}{(M+k)!}}, 
		\end{aligned}
		\end{equation}    
when $M\geq 2$, the inequality is always satisfied. For k=0, we have
    		\begin{equation}
		\begin{aligned}
		P^{\mu}_{M}(0)&=\sum_{l=0}^{\infty}\frac{(\mu)^{lM}e^{-\mu}}{(lM)!}
  \\ & \leq \sum_{l=0}^{\infty}\frac{(\mu)^{2l}e^{-\mu}}{(2l)!}
  \\ & = \frac{1+e^{-2\mu}}{2}.
		\end{aligned}
		\end{equation}
   For ${k=2}$, we have
      	\begin{equation}
		\begin{aligned}
		P^{\mu}_{M}(2)& \leq \sum_{l=1}^{\infty}\frac{(\mu)^{2l}e^{-\mu}}{(2l)!}
  \\ & = \frac{1}{2}(1+e^{-2\mu}-2e^{-\mu}).
		\end{aligned}
		\end{equation}
   For ${k=4}$
      		\begin{equation}
		\begin{aligned}
		P^{\mu}_{M}(4)& \leq \sum_{l=2}^{\infty}\frac{(\mu)^{2l}e^{-\mu}}{(2l)!}
  \\ & = \frac{1+e^{-2\mu}-2e^{-\mu}-\mu^{2}e^{-\mu}}{2}.
		\end{aligned}
		\end{equation}
  For ${k=6}$, we have
   		\begin{equation}
		\begin{aligned}
		P^{\mu}_{M}(6)& \leq \sum_{l=3}^{\infty}\frac{(\mu)^{2l}e^{-\mu}}{(2l)!}
  \\ & = \frac{1+e^{-2\mu}-2e^{-\mu}-\mu^{2}e^{-\mu}-2\mu^{4}e^{-\mu}/4!}{2}.
		\end{aligned}
		\end{equation}
Taking the formula above into Eq.~\eqref{devia}, we could get the deviation of even-photon components and total phase error rate with $M$ phase slices $E_{p}^{M}$.

\renewcommand{\theequation}{B\arabic{equation}}
		\begin{table*}[t]
			\caption{Experimental data.  }
			\setlength{\tabcolsep}{0.8cm}  
			\begin{tabular}
				{ccccccc}  \hline \hline
				Channel loss & \multicolumn{2}{c}{35 dB}& \multicolumn{2}{c}{40 dB} &\multicolumn{2}{c}{45 dB} \\ \hline
				$n_{\rm{det}}$ &\multicolumn{2}{c}{3701806} &\multicolumn{2}{c}{1196818} &\multicolumn{2}{c}{363094}   \\ 
   \cmidrule(r){2-3} \cmidrule(r){4-5} \cmidrule(r){6-7}
    \noalign{\smallskip}
			Detector &$D_{1}$ &$D_{2}$ & $D_{1}$ &$D_{2}$ & $D_{1}$ &$D_{2}$\\ 
				Detected 00 &46871 & 191 &15320 & 88 & 4669&40 \\
				Detected $\frac{\pi}{4}\frac{\pi}{4}$&49260 &93 &15749&49 &4835 & 42 \\ 
				Detected $\frac{\pi}{2}\frac{\pi}{2}$ &48173&71 &15602&42 &4716&45 \\ 
				Detected $\frac{3\pi}{4}\frac{3\pi}{4}$ &46155&156 &14682&74 &4573&42 \\ 
				Detected $\pi \pi$ &46633&180 &15041&82&4552&48  \\ 
				Detected $\frac{5\pi}{4}\frac{5\pi}{4}$ &45777&123 &14696&64&4448&53 \\ 
				Detected $\frac{3\pi}{2}\frac{3\pi}{2}$ &44793&84 &14432&53&4409&39  \\ 
				Detected $\frac{7\pi}{4}\frac{7\pi}{4}$ &51171&151 &16537&76&5029&57  \\ 
				Detected $0\pi$ & 160&72580&78&23467&44&6940  \\ 
				Detected $\frac{\pi}{4}\frac{5\pi}{4}$ &75&68516 &47&21991&28&6747 \\ 
				Detected $\frac{\pi}{2}\frac{3\pi}{2}$ & 65&69200&46& 22609&38&6681  \\ 
				Detected $\frac{3\pi}{4}\frac{7\pi}{4}$ & 159&70182&64&22637&34&6878 \\ 
				Detected $\pi 0$ &162&70496 &60&22976&35&6805 \\ 
				Detected $\frac{5\pi}{4}\frac{\pi}{4}$  &136& 73887 &67&23511&43&7292  \\
				Detected $\frac{3\pi}{2}\frac{\pi}{2}$  &89 &67507 &45&22036 &31&6539  \\ 
				Detected $\frac{7\pi}{4}\frac{3\pi}{4}$ &131&62112 &64&20205&30&6019 \\  \hline \hline
				
				\label{EXP_DET}
			\end{tabular}
		\end{table*}	
 \section{Phase error rate derivation details of discrete random phase}
For more clarity, we give the phase error rate derivation details of discrete random phase. As Eq.~\eqref{pe} shows, we give a discrete version of phase error rate
   \label{discrete}
\begin{equation}
\begin{aligned}
E_{p}^{M} & =\sum_{k=0}^{M/2-1} q_{2k}^{M} \\
& \leq \sum_{k=0}^{M/2-1} q_{2k} + \delta_{2k}\\
& \leq \sum_{k=0}^{\infty} q_{2k} + \sum_{k=0}^{M/2-1} \delta_{2k},
\end{aligned}
\end{equation}
the first inequality is obtained from Eq.~\eqref{devia}, and we expand the $q_{2k}$ sum range of the $k$ from $M/2-1$ to $\infty$ to get the second inequality. Combining with Eqs.~\eqref{pe} and \eqref{Ep}, the phase error rate can be written as
\begin{equation}
\begin{aligned}
E_{p}^{M} \leq \frac{e^{-\mu}\overline{Y}_{0}}{Q_{\mu}}+\frac{e^{-2\mu}+1-2e^{-\mu}}{2Q_{\mu}}+\sum_{k=0}^{M/2-1}\delta_{2k}.
\end{aligned}
\end{equation}

\renewcommand{\theequation}{C\arabic{equation}}

 \section{Utilizing Kato inequality to defend against the coherent attacks}

\label{Kato}
With the analysis of deviation between the continuous and discrete random phase~\cite{zeng2020sym}, our protocol applies only to the collective attacks for Eq.~\eqref{A1}. Here, we incorporate the Kato inequality~\cite{kato2020concentration}, which allows us to defend against coherent attacks for dependent random variables. The Kato inequality offers a tighter bound compared to the commonly used Azuma inequality~\cite{curras2021tight}.

Let us consider a sequence of Bernoulli random variables denoted as $\chi_{1},...,\chi_{n}$, and define $\Lambda_j$ as the sum of these random variables, i.e., $\Lambda_j=\sum_{u=1}^{j}\chi_{u}$. We also introduce $\mathcal{F}j$ as the $\sigma$-algebra generated by ${\chi_{1},...,\chi_{n}}$, representing the natural filtration of these Bernoulli random variables. Furthermore, let $\epsilon_{\rm{Ka}}$ represent the failure probabilities associated with the Kato inequality bound for sums of dependent random variables. By utilizing the findings presented in Refs.~\cite{curras2021tight,kato2020concentration}, we could find that
\begin{equation}
	\begin{aligned}
		{\rm Pr}\left [\sum \limits_{u=1}^{n}{\rm Pr}(\xi_{u}=1|\mathcal{F}_{u-1})-\Lambda_n \ge (b+a\left(\frac{2\Lambda_n}{n}-1\right))\sqrt{n}\right ]\\
  \leq {\rm exp}\left[\frac{-2(b^2-a^{2})}{1+\frac{4a}{3\sqrt{n}}}\right],\\
	\end{aligned}
	\label{concen1}
\end{equation}
by equating the right-hand sides of Eq.~(\ref{concen1}) to $\epsilon_{\rm{Ka}}$ and solving for $a$ and $b$, a tighter bound was derived~\cite{curras2021tight}. This improved bound can be formulated as
\begin{small}
\begin{equation}
	\begin{aligned}
		\sum \limits_{u=1}^{n}{\rm Pr}(\chi_{u}=1|\chi_{1},...,\chi_{u-1}) \leq \Lambda_n + \Delta_{\rm{Ka}}(n, \Lambda_{n}, \epsilon_{\rm{Ka}}),\\
	\end{aligned}
	\label{concen2}
\end{equation}
\end{small}where $\Delta_{\rm{Ka}}(n, \Lambda_{n}, \epsilon_{\rm{Ka}}) = \left[b+a(\frac{2\Lambda_{n}}{n}-1)  \right]\sqrt{n}$, where the $b$ and $a$ is set as
\begin{footnotesize}
\begin{equation}
\begin{aligned}
&a=\frac{3\left[72\sqrt{n}\Lambda_{n}(n-\Lambda_{n}){\rm ln} \epsilon_{\rm{Ka}} -16(n)^{3/2}{\rm ln}^{2}\epsilon_{\rm{Ka}}+9\sqrt{2}(n-2\Lambda_{n})a_{1}\right]} 
{4(9n-8{\rm ln} \epsilon_{\rm{Ka}})[9\Lambda_{n}(n-\Lambda_{n})-2n{\rm ln} \epsilon_{\rm{Ka}}]},\\
&b=\frac{\sqrt{18a^{2}n-(16a^{2}+24a\sqrt{n}+9n){\rm ln}\epsilon_{\rm{Ka}}}}{3\sqrt{2n}},
\end{aligned}
\end{equation}
\end{footnotesize}where $a_{1} = \sqrt{-n^{2}{\rm ln} \epsilon_{\rm{Ka}} [9\Lambda_{n}(n-\Lambda_{n})-2n{\rm ln} \epsilon_{\rm{Ka}}]}$,
and $\epsilon_{\rm{Ka}}$ represents the maximum failure probability among the bounds mentioned in Eq.~(\ref{concen2}). To estimate the upper bound of the phase error rate using the Kato inequality, we need to make a prediction of $\Lambda_n$, denoted as $\overline{\Lambda}_n$. This prediction is obtained during our security analysis process.

Considering the Kato inequality and the former phase error rate with finite-size analysis we obtain, the final phase error rate can be given by
\begin{equation}
	\begin{aligned}
\overline{E}_{p}^{M}\leq\frac{n_{\mu}E_{p}^{M}+\Delta_{\rm{Ka}}(n_{\mu}, n_{\mu}E_{p}^{M}, \epsilon_{\rm{Ka}})}{n_{\mu}}.
	\end{aligned}
\end{equation}

\section{Experimental details}
\label{exper details}
		\begin{table}[t]
			\caption{Efficiencies of devices in the measurement station.}
			\setlength{\tabcolsep}{1.2cm}  
			\begin{tabular}
				{cc}  \hline \hline
				Optical devices &  Attenuation \\ \hline
				Cir 2$\rightarrow$3  & 0.77 dB \\ 
				BS-3-1  & 3.61 dB \\ 
				BS-3-2  & 3.58 dB \\ 
				BS-4-1  & 3.80 dB \\ 
				BS-4-2  & 3.81 dB \\ 
				$PC_{1}$  & 0.18 dB \\ 
				$PC_{2}$  & 0.16 dB \\ \hline \hline
				\label{EFF_OPT}
			\end{tabular}
			
		\end{table}

The experimental results are summarized in Table~\ref{EXP_DET}, including the number of all detection events $n_{\rm{det}}$ and the number of detection events under different added phases. We denote the number of detection events under different added phases as ``Detected AB", where  ``A” (``B”) means that adding an A (B) phase to the pulse by Alice (Bob). The optical transmittance of the elements at Charlie's site are listed in Table~\ref{EFF_OPT}. The elements include the PM, PCs, Cir, and BS. The results of each channel are given accordingly. From the elements, we can obtain the proper additional loss to reach the total channel loss we need.


%

	\end{document}